\documentclass{article}

\begin{document}

\begin{center}
{\Large Stable and chaotic solutions of the complex Ginzburg-Landau equation
with periodic boundary conditions}

Jacob Scheuer$^{(1)}$ and Boris A. Malomed$^{(2)}$

$^{(1)}$ Department of Electrical Engineering, Technion, Haifa 32000, Israel

$^{(2)}$ Department of Interdisciplinary Studies, Faculty of Engineering,
Tel Aviv University, Tel Aviv 69978, Israel

Abstract
\end{center}

We study, analytically and numerically, the dynamical behavior of the
solutions of the complex Ginzburg-Landau equation with diffraction but
without diffusion, which governs the spatial evolution of the field in an
active nonlinear laser cavity. Accordingly, the solutions are subject to
periodic boundary conditions. The analysis reveals regions of stable
stationary solutions in the model's parameter space, and a wide range of
oscillatory and chaotic behaviors. Close to the first bifurcation
destabilizing the spatially uniform solution, a stationary single-humped
solution is found in an asymptotic analytical form, which turns out to be in
very good agreement with the numerical results. Simulations reveal a series
of stable stationary multi-humped solutions.

\section{INTRODUCTION}

There is a great current interest in dynamics described by the complex
Ginzburg-Landau equation (CGLE) [1], which is a generic nonlinear model with
various physical applications, including binary-fluid thermal convection
[2], semiconductor lasers [3], nonlinear optical fibers (see Ref. [4] and
references therein), etc. In most applications, the boundary conditions (BC)
supplementing the CGLE are an inherent type of the underlying physical
model. In particular, in the case of the thermal convection and laser
cavities, the simplest and most fundamental type of BC are the periodic
conditions, as they make it possible to study the intrinsic dynamics of CGLE
avoiding obscuring effects of wave reflection from edges.

It should be stressed that the dynamics of the CGLE with periodic BC is
principally different from that of an infinite system, as various solutions
(even chaotic) which are stable under periodic BC may be easily destabilized
in a much longer system [5]. Another important difference from the infinite
limit is that the equation subject to periodic BC is equivalent to an
infinite chain of evolution equations for amplitudes of the corresponding
Fourier expansion. Using this approach, it has been proved that a
finite-dimensional attractor exists in the system's phase space [6] (which
is true even in the case of two-dimensional CGLE [7]). Parallel to this,
different technical versions of an approximation based on a finite-mode
(Galerkin) truncation of the above-mentioned infinite chain of equations
were developed, helping to understand the onset of dynamical chaos in the
CGLE [8]. Although it was argued that the Galerkin truncation does not
always describe adequately the transition to chaos in CGLE [9], it has been
demonstrated that, generally, the finite-mode truncation of the CGLE with
periodic BC provides for quite accurate results [10].

For some particular cases, a chain of bifurcations leading to chaos in CGLE
under periodic BC was found by means of direct numerical simulations [11]:
the chain features transitions limit-cycle $\rightarrow $ two-torus $%
\rightarrow $ three-torus $\rightarrow $ chaos, $n$-torus being a
quasi-periodic dynamical regime with $n$ incommensurate frequencies. Many
other nonlinear modes are unstable at onset (i.e., when they appear), but
they may be later restabilized through higher-order bifurcations [12].

In this paper, we study analytically and numerically the dynamical behavior
and, in particular, stable solutions of the cubic CGLE, subjected to
periodic BC, without diffusion, while diffraction (which formally
corresponds to an imaginary part of the diffusion coefficient) is present.
This situation is typical for laser cavities and other nonlinear optical
problems in the spatial domain (see Ref. [13] and [14]), as light beams are
subject to transverse diffraction but not diffusion.

A general form of the one-dimensional CGLE with cubic nonlinearity and
without diffusion is 
\begin{equation}
iu_{t}+\frac{1}{2}u_{xx}\left( 1+iR\right) \left| u\right| ^{2}u=iu\,.
\end{equation}
Here, $u=u(x,t)$ is the complex field which depends on the time $t$ and the
spatial coordinate $x$, $R$ is a real positive parameter that represents
cubic losses and/or gain saturation. The coefficients of the spatial
dispersion, the Kerr nonlinearity and the linear gain (the latter one
accounts for the term on the right-hand side of Eq. (1)) are all normalized
to 1. In nonlinear optics, where Eq. (1) finds its most important
application, it describes the spatial evolution of the electromagnetic field
envelope $u(t,x)$ in a planar waveguide or laser cavity with linear gain,
cubic loss and Kerr nonlinearity. In this case, the evolutional variable $t$
is, in fact, not time but the propagation distance (which is frequently
designated $z$), and the second term in Eq. (1) accounts for the transverse
diffraction [13].

Equation (1) also describes the temporal or spatial evolution of the
electric field in a (1+1)-dimensional semiconductor laser medium with gain
saturation. The second term in Eq. (1) then accounts for the diffraction in
space or dispersion in the temporal domain, and the cubic conservative term
represents a normalized anti-guiding or linewidth-enhancement factor.

In either case, we supplement Eq. (1) with periodic BC, , where $L$ is the
system's length (in the transverse direction, in the case of the planar
waveguide). In its literal form, Eq. (1) with periodic BC applies to a
planar waveguide closed into a cylindrical surface. Such a configuration can
be easily implemented in Vertical Cavity Surface-Emitting Lasers (VCSEL)
[3]. However, the most widespread application of Eq. (1) subject to periodic
BC is the description of a ring cavity laser, in which some ingredients of
the system (e.g., the saturable amplifier and nonlinear waveguide) may be
separated. In this context, Eq. (1) has the meaning of an average master
equation that govern the evolution of the electromagnetic field in the
cavity. Thus, the model is controlled by two dimensionless parameters, $R$
and $L$. Despite a number of results obtained in the above-mentioned works
for CGLE of a more general type with periodic BC, the case of zero diffusion
is fundamentally important for applications to nonlinear optics and must be
considered separately.

Equation (1) has an obvious continuous-wave (CW), i.e., stationary spatially
uniform solution, 
\begin{equation}
u_{0}=R^{-1/2}\exp \left( iR^{-1}t\right) \,.
\end{equation}

A straightforward stability analysis, taking into account the periodic BC,
shows that the CW solution (2) is unstable to small perturbations if the
length $L$ exceeds a critical value: 
\begin{equation}
L_{\mathrm{cr}}=\pi \sqrt{R}.
\end{equation}

A primary objective of this work is to study the model's dynamics in the
case $L>L_{\mathrm{cr}}$. To this end, we solved Eq. (1) numerically, using
the uniform solution (2) perturbed by a small-amplitude harmonic with the
fundamental wave-number ($2\pi /L$) as initial conditions. We have found
that the long-time evolution of the solutions exhibits a wide variety of
behaviors (stable, oscillating, chaotic), depending on values of the
parameters $R$ and $L$.

The rest of the paper is organized as follows. In section 2, we display a
phase diagram, which summarizes the dynamical behavior of the system in
terms of the ($R,L$) parametric plane. In this section, we also find, in an
asymptotic analytical form, the first spatially nonuniform stable stationary
solution which sets in after the stability loss of (2), and compare it to
numerical results. Although the procedure leading to the analytical solution
is quite simple, the solution itself appears to be a new one. In section 3,
we study higher-order stable stationary solutions which have a multi-humped
structure. In section 4, we study various oscillatory dynamical regimes and
section 5 concludes the paper.

All these dynamical regimes admit straightforward interpretation in terms of
field patters in the laser cavity which is modeled by Eq. (1). In fact, as
it was already mentioned above, we study in detail two vast classes of
regular stable solutions: stationary and oscillatory (we also find chaotic
dynamical regimes, but, from the viewpoint of applications, they are
interesting mainly in the sense of choosing parameter values in order to
avoid chaos). The physical relevance of both types of regular dynamical
regimes is quite obvious. In particular, we will find different parametric
regions that feature stable stationary states with one or several humps
(local maxima). This finding has straightforward applications in term of the
laser cavity, making it possible, for instance, to simultaneously generate
several output light beams. Stable oscillatory regimes are relevant too, as
they can help the laser to generate a temporally modulated (pulsed) beam
with various types of the modulation, depending on the quasi-harmonic or
strongly anharmonic character of the oscillations (we find stable solutions
of both types).

\section{A DIAGRAM OF DYNAMICAL REGIMES AND BASIC STATIONARY STATES}

Figure 1 displays a diagram of different dynamical regimes in the parametric
plane ($L,R$) generated by numerical simulations of Eq. (1). The figure
shows several regions of regular (stationary or oscillatory) solutions,
which are separated by chaotic layers. Regular solutions existing in
different regions exhibit different spatial patterns. Each region is divided
into sub-regions of stationary and oscillating solutions.

In region I, which is , the uniform solution (2) is stable. Direct
simulations demonstrate that, indeed, there is no other attractor in this
region. The first stable stationary non-uniform solution found beyond the
boundary (3) is shown in Fig. 2. Close to the boundary, this solution can be
approximated analytically as a combination of a uniform field and the first
two spatial harmonics: 
\begin{equation}
u(x,t)=R^{-1/2}\exp \left( iR^{-1}t-i\omega _{2}t\right) \cdot \left[
1+a_{0}+a_{1}\cos \left( kx\right) +a_{2}\cos \left( 2kx\right) \right]
\end{equation}
where $k=2\pi /L$, $a_{1}$, $a_{0}$ and $a_{2}$ are, respectively, small
amplitudes of the fundamental, zeroth, and second harmonics,\ $\omega _{2}$
is a frequency shift, and it is expected that $a_{0}$, $a_{2}$ and $\omega
_{2}$ are $\,\sim a_{1}^{2}$.

To proceed with the analytical consideration of the solution (4), we
eliminate $a_{0}$, $a_{2}$ and $\omega _{2}$, substituting Eq. (4) into Eq.
(1) and equating terms at order $a_{1}^{2}$. We notice that the ansatz (4)
allows one to assume that $a_{0}$ is purely real. Obviously, is real too,
while $a_{1}$ and $a_{2}$ may be complex: 
\begin{equation}
a_{1}\equiv a_{11}+ia_{12},\,a_{2}\equiv a_{21}+ia_{22}\,.
\end{equation}
Substituting (5) into (4) and solving for both real and imaginary parts, we
obtain the following expressions at order $a_{1}^{2}$: 
\begin{eqnarray}
\omega _{2} &=&\left( 1+\frac{1}{R^{2}}\right) a_{11}a_{12},\,a_{0}=\frac{3}{%
4}a_{11}^{2}-\frac{1}{4}a_{12}^{2}-\frac{1}{2R}a_{11}a_{12}\,,  \nonumber \\
a_{21} &=&\frac{1}{4}a_{11}^{2}+\frac{1}{12}a_{12}^{2}-\frac{R}{6}%
a_{11}a_{12}\,,\,\,a_{22}=\frac{R}{4}a_{11}^{2}+\frac{R}{12}%
a_{12}^{2}-\left( \frac{R^{2}}{24}-\frac{1}{8}\right) a_{11}a_{12}\,.
\end{eqnarray}

Next, we equate coefficients in front of the fundamental harmonic (obtained
after the substitution of (4) into (1)) to find the amplitude $a_{1}$ at
order $a_{1}^{3}$. It is convenient to introduce a small positive parameter $%
\delta $ (overcriticality) which measures proximity of the system to the
bifurcation point (3), 
\begin{equation}
\delta \equiv 4R^{-1}-k^{2}
\end{equation}
(recall $k=2\pi /L$). The amplitude a1 can then be found as a function of $%
\delta $, 
\begin{equation}
a_{12}=Ra_{11},\,\,a_{11}^{2}=\frac{3R}{R^{4}-9R^{2}+30}\delta \,.
\end{equation}
Figures 3 and 4 present comparison between the analytical solution (4) - (8)
and a numerical one for different values of $R$ and $L$. Up to relatively
large values of the overcriticality ($\delta =0.30$ in Fig. 3), fairly good
agreement between the numerical and analytical solutions is evident. A very
good agreement was also found between the analytically calculated frequency
shift and the shift that was extracted from numerical simulations. With
further increase of $\delta$, a deviation between the analytical and the
numerical solutions becomes visible (at $\delta \approx 0.40$ in Fig. 4).

The parameter range, in which this stationary solution is stable, is a part
of region II in Fig. 1. Although the region is narrow, it has an internal
structure, being divided into three subregions, hosting the above stationary
solution and oscillatory ones. A detailed consideration of the subregions is
given in section 4. Above the region II (at larger $L$ and $R<2.6$), the
simulations reveal solutions which are chaotic in time.

\section{HIGHER-ORDER STABLE STATIONARY SOLUTIONS}

Another region partly filled with stable stationary solutions is found as $L$
is increased, namely, region III in Fig. 1. In contrary to the solutions
considered above, the profile of the solution in the region III, a typical
example of which is depicted in Fig. 5, has two maxima (with different
values) and two minima, being similar to solutions discussed, in the
framework of a similar model, in Ref. [15]. In this region, the amplitude of
the second harmonic is not small in comparison with that of the fundamental
harmonic, giving rise to the double-humped profile of the solution. A
minimum value of the nonlinear-loss coefficient for which region III exists
is $R=1.06$, the corresponding value of the system's length being $L=5.9$.
Regions II and III merge at $R>2.6$ and form a single region (see Fig. 1).
The dotted line inside the united II - III region in Fig. 1 indicates a
border between the above-mentioned single-humped and double-humped
stationary solutions. This border is a direct continuation of the upper
border of region II at lower values of $R$, strongly suggesting that the
transition between the single- and double-humped patterns is caused by
destabilization of the single-humped solution. In fact, region III, as well
as II, is also divided into several subregions exhibiting stationary and
oscillating regular solutions.

Above region III, there is a well-defined stripe in which the behavior of
the system is chaotic. For higher values of $L$, still another range of
stable stationary solutions, region IV in Fig. 1, is revealed. Stationary
solutions in this region also have two lobes, like in region III, but its
second-harmonic component is larger, giving rise to a profile which is
similar to the solution discussed in Ref. [16] (a cnoidal-wave-type, see
Fig. 6). It should be noted, however, that region IV is rather narrow.

The next stability range for stationary solutions is found directly above
region IV, at $R>1.72$ and $L>11.8$ (region V in Fig. 1). In this region,
stable solutions have four maxima (two pairs of local maxima with of
different heights), see Fig. 7. It seems plausible that this solution was
obtained via a doubling bifurcation of the double-humped solution found
above in region II, especially since the minimal system length for this
region ($L=11.8$) is exactly twice the minimal length for region III ($L=5.9$%
).

This pattern of changing the structure of stationary solutions develops at
larger values of $L$, with stability regions featuring stationary solutions
that appear to be produced by tripling and quadrupling the basic stationary
solution found in region III (see Fig. 5). As $L$ is increased, reaching
region VI in Fig. 1, a new stable stationary solution of Eq. (1) appears,
characterized by local maxima with three different amplitudes, see Fig. 8.
The minimum value of $R$ for which this solution appears is $R=1.55$, at $%
L=18$. This solution evolves gradually from the four-lobed solution found in
region V and described above, see Fig. 7. As $L$ is increased, the two small
maxima move outwards; simultaneously, two additional humps evolve between
the small maxima and the central high maxima (see Fig. 9). Eventually, these
two humps evolve into two additional maxima.

\section{OSCILLATORY SOLUTIONS}

As it was mentioned before, each region containing stable regular solutions
is actually divided into several subregions that exhibit a wide variety of
dynamical behaviors. As an example, we present here a detailed description
of the intrinsic structure of region II.

Figure 10 illustrates the intrinsic structure of region II for $0.2<R<2.2$.
Although this region is relatively narrow, it is divided into three
subregions. In the first subregion (IIa), the established solution is always
the stationary one that was described in section 2. For larger values of $L$
(subregion IIb), the solution does not remain stationary; instead, it
becomes an oscillatory function of $t$ (see Fig. 11).

Further increase of $L$ results in a change of the character of the
oscillations in the subregion IIc: while the oscillations in the subregion
IIb were quasi-harmonic, in the subregion IIc they acquire a more complex
anharmonic structure. Clear characterization of the oscillations of
different types is provided by examining the corresponding time dependence
of the average field power, which is defined in a straightforward way, 
\begin{equation}
P(t)=L^{-1}\int_{0}^{L}\left| u(x,t)\right| ^{2}dx,
\end{equation}
and also by the temporal Fourier transform (power spectrum) of the
time-dependent power (9). In particular, in Fig. 12 it is easy to see
distinction between the aforementioned quasi-harmonic and strongly
anharmonic oscillations, in terms of the dependences $P(t)$.

Figure 13 shows the power spectrum of $P$ obtained at $R=0.7$, while varying 
$L $ between $L=2.9$ and beyond the upper border of region II ($L=3.15$). In
this figure, the dc (zero-harmonic) component of the Fourier transform was
removed in order to stress the dynamical behavior. As $L$ crosses the upper
border of region IIa, two symmetric harmonics with opposite frequencies
appear, indicating that the solution is oscillating quasi-harmonically. As $%
L $ is increased further, the frequency of the oscillations drops to a
minimum value, after which several higher-order harmonics appear, and the
dominant frequency increases with $L$. At the point with the lowest value of
the oscillation frequency, a transition between two different dynamical
regimes, viz., quasi-harmonic oscillations whose frequency decreases with $L$
and complex multi-frequency oscillations whose dominant frequency increases
with $L$, takes place.

Further increase of $L$ eventually results in a transition to dynamical
chaos. A similar behavior is observed when decreasing $R$ at constant $L$.

It should be emphasized that all the other stability regions marked in Fig.
1 are also divided into subregions which exhibit various oscillatory
dynamical regimes, see examples shown in Fig. 14 for region V, and in Fig.
15 for region VI. However, a detailed analysis of these subregions is beyond
the scope of the present paper.

\section{CONCLUSIONS}

We have studied the dynamical properties of the solutions of the complex
cubic Ginzburg-Landau equation without diffusion, which is subject to
periodic boundary conditions. This equation is a fundamental model of
nonlinear-optical cavities characterized by saturable gain. Systematic
investigation of its dynamics is necessary, as it is essentially different
from both a more general Ginzburg-Landau equation with its entire
coefficients complex (i.e., including also diffusion), and from its
infinite-length counterpart.

We found that the model exhibits a complex structure, including regions of
stable stationary, oscillatory, and chaotic solutions in its parameter
plane. For a case of strong nonlinear loss, the system features a
multi-layered structure in the parameter plane, with a different dynamical
behavior in each layer. A boundary between adjacent layers implies the
existence of a bifurcation destabilizing the solution in one of layers. This
could be easily seen by examining the boundary between regions II and III in
Fig. 1, i.e., the upper boundary of region II at $R>2.6$: it is a direct
continuation of the upper boundary of region II which, at $R<2.6$, separates
it from a chaotic zone. It should be stressed that we did not observe
hysteresis (overlapping) between stable solutions of different types.

We have found an asymptotic analytical single-humped solution for the case
when the system is close to the first bifurcation destabilizing the
spatially uniform CW solution. Even when the system is not really close to
the bifurcation point, the analytical solution is found to be in good
agreement with numerical results. Numerically, we have also found a series
of stable stationary solutions of a multi-humped type.

One of the authors (B.A.M.) appreciates a grant ''Window-on-Science'',
provided by the European Office for Research and Development of the US Air
Force.

\begin{center}
{\Large Figure Captions}
\end{center}

Fig. 1. The dynamical phase diagram of the system, showing regions of
regular and chaotic behavior in the parameter plane ($R,L$), see detailed
description of the regions in the text.

Fig. 2. A stable stationary nonuniform solution found for $R=1.0$ and $L=3.4$
(inside region II of Fig. 1), which has a single maximum and a single
minimum. In this figure and in Figs. 3 - 9 below, shown is $|u(x)|$ in the
interval $0<x<L$.

Fig. 3. Comparison between the analytical (circles) and the numerical
(solid) stationary solutions for $R=0.7$ and $L=2.7$. At these values of $R$
and $L$, Eq. (7) gives the value of the overcriticality parameter $\delta
=0.30$, and Eq. (3) yields the critical value beyond which the nontrivial
stationary solution may exist, $L_{\mathrm{cr}}=2.628$.

Fig. 4. Comparison between the analytical (dashed) and numerical (solid)
stationary solutions for $R=1.4$ and $L=4.01$. At this value of $R$, one has 
$L_{\mathrm{cr}}=3.717$ and $\delta $ = 0.40.

Fig. 5. A typical profile of the stable stationary solution found in region
III, for $R=2.0$ and $L=6.0$. The solution exhibits two maxima and two
minima.

Fig. 6. The profile of the stable stationary solution found solution in
region IV for $R=3.2$ and $L=11.8$, featuring a cnoidal-wave-like solution
with two maxima.

Fig. 7. The profile of the stable stationary solution found in region V for $%
R=2.5$ and $L=12.0$, which exhibits a double-period solution having local
maxima and minima with different amplitudes.

Fig. 8. The profile of the stable stationary solution obtained in region VI
for $R=1.8$ and $L=18.8$, with three maxima having different amplitudes.

Fig. 9. The evolution of the stable stationary solution observed when
crossing from region V into VI, for $R=2.0$ and $L=13.2$ (solid), $17.0$
(dotted), $19.0$ (dashed), $21.0$ (dashed-dotted).

Fig. 10. The inner structure of region II which reveals three subregions
that feature, respectively, stable stationary patterns and quasi-harmonic or
strongly anharmonic oscillatory solutions, as it is described in the text in
detail.

Fig. 11. The oscillatory solution as a function of $t$, found in region IIb
(for $R=0.3$ and $L=1.85$).

Fig. 12. Comparison between the time dependence of the average field power $%
P(t)$ in regions IIb (for $R=0.7$ and $L=2.97$, solid line) and IIc (for $%
R=0.7$ and $L=3.02$, dashed line).

Fig. 13. The evolution of the power spectrum of $P(t)$ inside the interval $%
2.9<L<3.15$ at fixed $R=0.7$, indicating at the existence of two different
types of oscillatory regimes.

Fig. 14. Examples of the temporal evolution of the field corresponding to
different types of oscillatory dynamical regimes in region V: (a) $R=4.0$, $%
L=17.5$; (b) $R=3.0$, $L=17.4$.

Fig. 15. Field evolution in time in region VI, for $R=1.7$ and $L=16.1$.

\end{document}